\documentclass[12pt,preprint]{aastex}

\usepackage{CJK}

\usepackage{epsfig}

\shorttitle{organics in the solar system}
\shortauthors{Kwok}

\begin{document}

\title{Organics in the Solar System}


\author{Sun Kwok\altaffilmark{1, 2}}

\affil{$^1$ Laboratory for Space Research,  The University of Hong Kong, Hong Kong, China\\
$^2$ Department of Earth, Ocean, and Atmospheric Sciences, University of British Columbia, Vancouver, B.C., Canada}
 \email{sunkwok@hku.hk; skwok@eoas.ubc.ca}

\begin{abstract}

Complex organics are now commonly found in meteorites, comets, asteroids, planetary satellites, and interplanetary dust particles.  The chemical composition and possible origin of these organics are presented.  Specifically, we discuss the possible link between Solar System organics and the complex organics synthesized during the late stages of stellar evolution.  Implications of extraterrestrial organics on the origin of life on Earth and the possibility of existence of primordial organics on Earth are also discussed.

\end{abstract}


\maketitle

\vspace{3in}

\noindent{Invited review presented at the International Symposium on Lunar and Planetary Science, accepted for publication in Research in Astronomy and Astrophysics.}
\clearpage

\section{Introduction}

In the traditional picture of the Solar System, planets, asteroids, comets and planetary satellites were formed from a well-mixed primordial nebula of chemically and isotopically uniform composition.  The primordial solar nebula was believed to be initially composed of only atomic elements synthesized by previous generations of stars, and current Solar System objects later condensed out of this homogeneous gaseous nebula.  Gas, ice, metals, and minerals were assumed to be the primarily constituents of planetary bodies \citep{suess}.

Although the presence of organics in meteorites was hinted as early as the 19th century \citep{berzelius}, the first definite evidence for the presence of extraterrestrial organic matter in the Solar System was the discovery of paraffins in the Orgueil meteorites \citep{nagy}.  Since it was common belief at that time that organic matter resides in the sole domain of the Earth, it was speculated that the Orgueil meteorite may have originated from Earth and returned after millions of years of travel in interplanetary space \citep{bernal}.
The current acceptance of the existence of extraterrestrial organics only came after the identification of amino acids, hydrocarbons, and aromatic/aliphatic compounds in the Murchison meteorite \citep{kven, cronin}.

With modern spacecrafts, we now have the capability of flying instruments to Solar System objects to make in-situ measurements, and to return sample to Earth for further analysis.  With a variety of laboratory techniques at our disposal, extraterrestrial organic compounds can be detected and identified with a great degree of certainty. 

\section{Organic matter on Earth}

Although biomass may first seem to be the major reservoir of organics on Earth, in fact the great majority of organics on Earth is in the form of kerogen, a macro-organic substance found in sedimentary rocks \citep{falkowski2000} (Table 1).  Kerogen is amorphous in structure and is made of random arrays of aromatic carbon sites and aliphatic chains with functional groups.  
Kerogen originated from remnants of past life, basically algae, land plants, and animals.  Under high temperature and pressure, kerogen gradually dissociated into petroleum and natural gas \citep{vanden}.  
Except for small amounts of methane in hydrothermal vents \citep{sherwood, taran}, the majority of organics on Earth is biological in origin.

\begin{table}[h]
 \caption{Organic carbon pools in the major reservoirs on Earth$^a$\label{falkowski}
}
    \begin{tabular}{lr}
\hline
\noalign{\vskip6pt}
\multicolumn{1}{c}{Pools} & {Quantity (Gt)}\\
\noalign{\vskip6pt}
\hline
Oceans&1,000\\
Lithosphere\\
\hspace{10pt}Kerogens &15,000,000\\
Terrestrial biosphere&\\
\hspace{10pt}Living biomass &600-1,000$^b$\\
\hspace{10pt}Dead biomass &1,200\\
Aquatic biosphere &1-2\\
Fossil fuels &\\
\hspace{10pt}Coal &3,510\\
\hspace{10pt}Oil &230\\
\hspace{10pt}Gas &140\\
\hspace{10pt}Other (peat)&250\\
\noalign{\vskip6pt}
\hline
\end{tabular}
\tablenotetext{a}{Table adapted from \citet{falkowski2000}.}
\tablenotetext{b}{Additional 50\% may be in the form of subsurface bacteria and archaea \citep{whitman}.}
\end{table}

\section{Organic matter in the Solar System}

Although not initially anticipated, organics are now found in all classes of Solar System objects.  A brief summary is given below.

\subsection{Meteorites}

Organics in meteorites are primarily found in carbonaceous chondrites, a rare class of meteorites that are believed to be most pristine of all meteorites.  Through the use of different solvents, various components of the meteorite can be extracted and analyzed.  Modern analysis has found that almost all basic biologically relevant organic molecules are present in carbonaceous meteorites.
Organics identified in the soluble component of carbonaceous chondrites include carboxylic acids, sulfonic and phosphonic acids, amino acids, aromatic hydrocarbons, heterocyclic compounds, aliphatic hydrocarbons, amines and amides, alcohols, aldehydes, ketones, and sugar related compounds \citep{remusat}.  Altogether over 14,000 compounds with millions of diverse isomeric structures have been found  \citep{schmitt_kopplin2010}.

Specifically, amino acids of equal mixture of D and L chirality and non-protein amino acids not found in the biosphere have been identified in carbonaceous chondrites.  Unusual nucleobases (e.g., 2,6-diaminopurine, and 6,8-diaminopurine) beyond the 5 used in terrestrial biochemistry (adenine (A), cytosine (C), guanine (G), thymine (T), and uracil (U)) are also found \citep{callahan}.  Many more amino acids have been identified in meteorites than those that are used in our terrestrial biochemistry \citep{Pizzarello2017}.  
While these molecules are among the large family of possible organic molecules  \citep{cleaves}, it is clear that these biomolecules are not related to terrestrial life.
The decreasing molecular abundance with increasing carbon number within the same class of compounds also provide additional evidence for their abiotic origin.

The large fraction (70--90\%) of organic carbon in carbonaceous chondrites is in the form of a complex, insoluble, macromolecular material often referred to as insoluble organic matter \citep[IOM,][]{cronin}.
IOM has been analyzed by both destructive (thermal and chemical degradations followed by gas chromatography–mass spectrometry) and nondestrutive (nuclear magnetic resonance,  fourier transform infrared spectroscopy, x-ray near-edge spectroscopy,  electron paramagnetic resonance, high-resolution transmission electron microscopy) means, yielding a chemical structure consisting of small islands of aromatic rings connected by short aliphatic chains \citep{cody2002, cody2011}.  Isotope anomalies in IOM suggest that it is probably of interstellar origin \citep{busemann}.

The exact relationship between the soluble and insoluble components of organics is not clear, although that there is evidence some of the soluble components could be released from the IOM through hydrothermal alteration \citep{sephton1998, yabuta}.

\subsection{Comets}

Comets are no longer believed to be just ``dirty ice balls'' but contain a significant amount of organics in the nuclei \citep{sandford, cody2011}.  A large variety of gas-phase molecules from the volatile component of comets have been detected through their rotational/vibrational transitions by remote telescope observations  \citep{mumma2011}.    
The very high spectral resolution ($\sim 10^7$) of modern millimeter/submillimeter-wave spectroscopy allows for precise identification of molecular species, and organic molecules with as many as 10 atoms (ethylene glycol, HOCH$_2$CH$_2$OH) have been identified in comets.
The mass spectrometer aboard {\it Rosetta's Philae} lander has detected an array of organic compounds on the surface of the comet 67P/Churyumov-Gerasimenko \citep{goesmann}.  Among the molecules detected in the sample returned from Comet 81P/Wide 2 by the {\it Stardust} mission include the amino acid glycine, which has a carbon isotopic ratio suggestive of extra-solar origin \citep{elsila2009}.  

Macromolecular compounds similar to IOM in meteorites are also detected in cometary dust \citep{fray}.  Infrared spectra of this material show the 3.3 $\mu$m aromatic C--H stretch and 3.4 $\mu$m aliphatic C--H stretch, suggesting cometary dust contains both aromatic and aliphatic materials \citep{keller2006}.

\subsection{Planets and planetary satellites}

The recent discoveries of thiophenic, aromatic, and aliphatic compounds in drilled samples from Mars' Gale crater \citep{eigen} has generated a great deal of publicity in the popular press.  These organics can be traced to kerogen-like materials as their precursor.  The detection of organics in plumes from Enceladus suggests the presence of large reservoir of complex macromolecular organics in subsurface oceans \citep{postberg}.

Saturn's moon Titan has been found to have extensive lakes of liquid methane and ethane, as well as organic particles in surface dunes.   
Titan's atmosphere is filled with organic haze.  The 3.4 $\mu$m C--H stretching band from aliphatic hydrocarbons have been observed in the atmosphere of Titan by the Cassini/Visual Infrared Mapping Spectrometer instrument \citep{kim}.  
Radar imaging observations from {\it Cassini} have found hundreds of lakes and seas filled with liquid methane \citep{legall}. 
The total amount of organic carbon in Titan is estimated to be 360,000 Gt in atmosphere, 16,000--160,000 Gt in lakes, and 160,000--640,000 Gt in sand dunes \citep{lorenz}, which is much larger than the total amount of fossil fuels on Earth (Table 1).
  
The popular model for the chemical structure of these organic particles in Titan is tholins, an amorphous hydrogenated carbonaceous compound rich in nitrogen \citep{sagan1979}.

\subsection{Asteroids}

\begin{figure}
\centering
\includegraphics[width=0.6\textwidth]{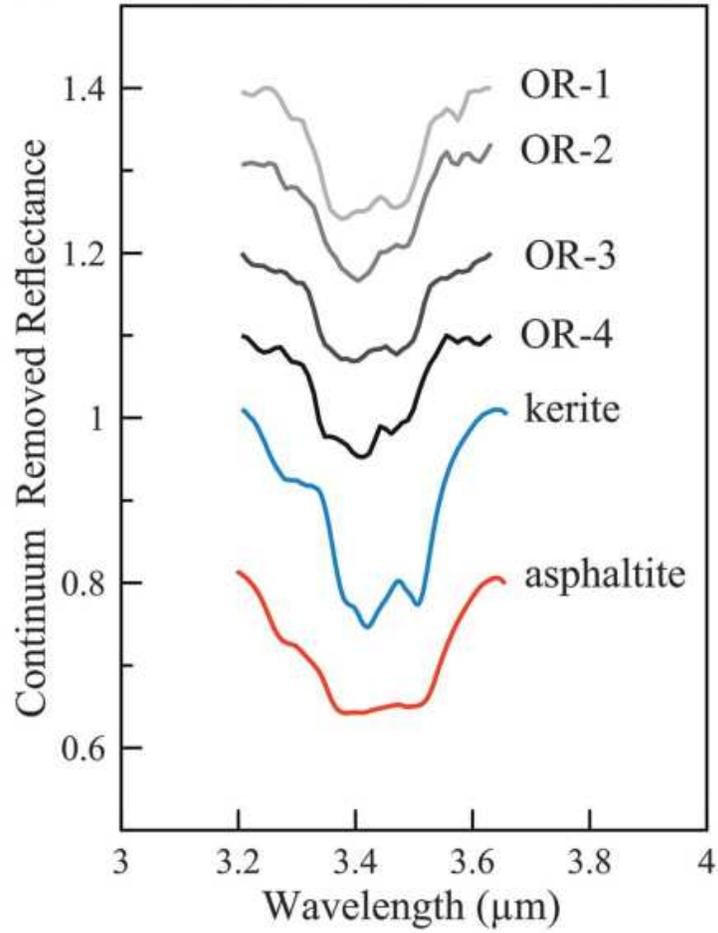}
\caption{The 3.4 $\mu$m aliphatic C--H stretching band is detected in several organic spots in the dwarf planet Ceres (OR-1 to OR-4).  Spectra of terrestrial Kerite and asphaltite are also shown for comparison.  Figure adapted from \citet{desanctis}.}
\label{ceres}
\end{figure}

The red color and low albedo of D-type asteroids has led to the hypothesis that they are covered by kerogen-like macromolecular organics on the surface \citep{dale}.   The recent detection by the Visible and InfraReed Mapping Spectrometer on board the {\it Dawn} spacecraft of the 3.4 $\mu$m aliphatic C--H stretch (Figure \ref{ceres}) over a  1000 km$^2$ area near the Ernutet Crater of the dwarf planet Ceres gives definite confirmation that complex organics are present in the main asteroid belt \citep{desanctis}.

\subsection{Micrometeorites and interplanetary dust particles}

Micrometeorites of sizes 20 $\mu$m to 1 mm can be collected in clean, isolated environment such as sand, deep sea sediments, Greenland lake sediments and Antarctic ice and snow.  Mircometeorites are  believed to be the dominant source of extraterrestrial matter currently being accreted by the Earth.  Some Antarctic micrometeorites have been found to contain large fraction of organics, including amino acids \citep{dobrica}.  The presence of complex organics in micrometeorites suggests that extraterrestrial organics can survive atmospheric passage and impact.  

\begin{figure}
\centering
\includegraphics[width=0.6\textwidth]{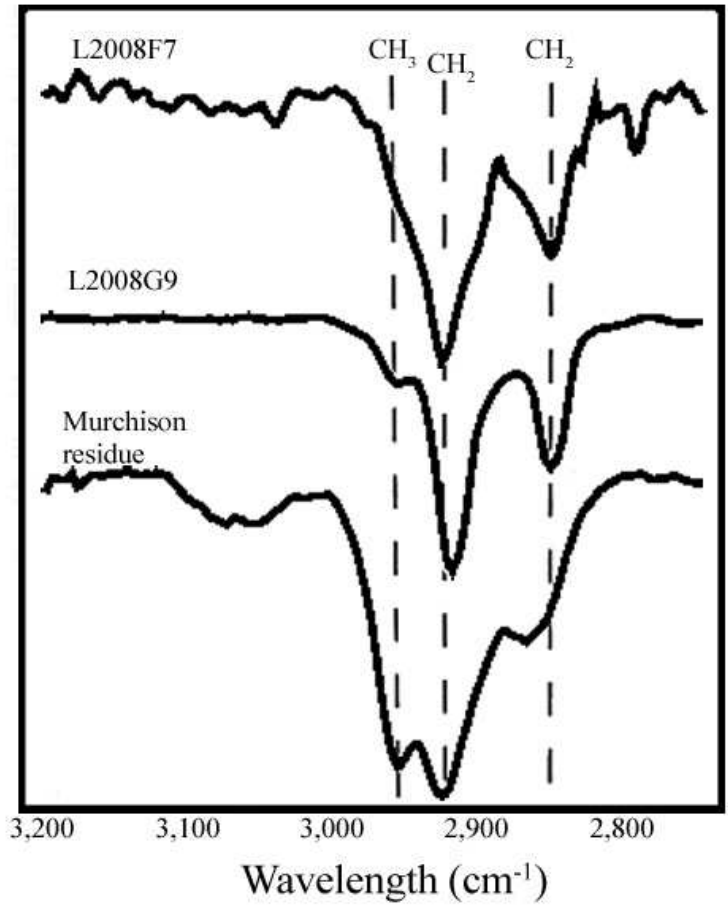}
\caption{The infrared absorption spectra of interplanetary dust particles L2008F7 and L2008G9 and the Murchison meteorite showing the aliphatic stretching modes of CH$_2$ and CH$_3$ near 3.4 $\mu$m (marked by vertical dashed lines). Figure adapted from \citet{flynn03}.}
\label{flynn}
\end{figure}

Laboratory analysis of interplanetary dust particles (IDP, sizes 1--30 $\mu$m) collected in the Earth's stratosphere show definite evidence of complex organics \citep{flynn03}.  Figure \ref{flynn} shows the 3.4 $\mu$m feature of aliphatic hydrocarbon in the infrared spectra of two IDPs and the Murchison meteorite.  Since IDPs are believed to have originated from comets and asteroids, this gives us indirect evidence that similar organics are present in comets and asteroids.  The anomalous H and N isotopic ratios of IDP organics suggest that they are of presolar origin \citep{keller2004}. 

\subsection{Outer solar system objects}

\begin{figure}[h]
\begin{center}
\includegraphics[width=3.0in]{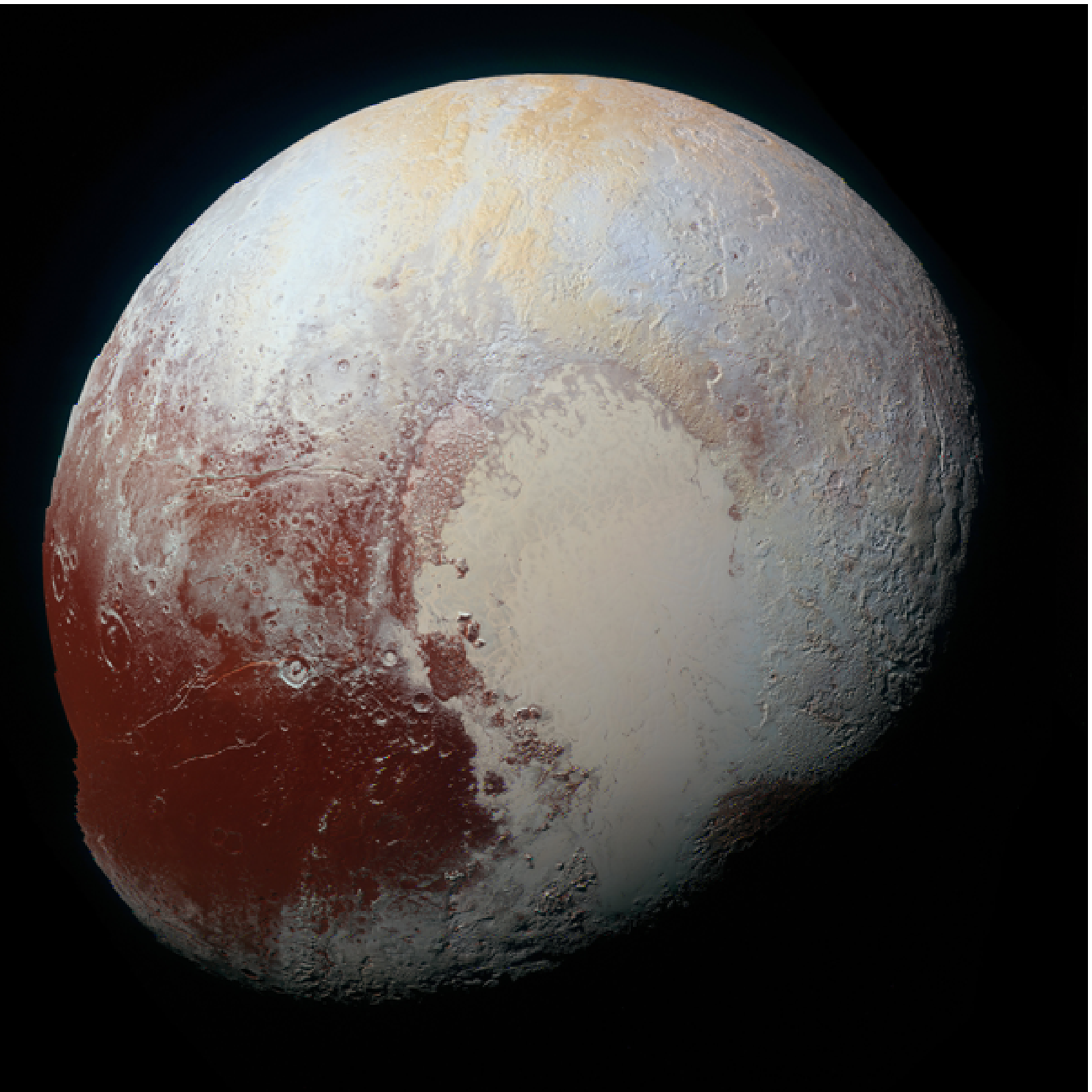} 
 \caption{An image of Pluto as observed by the {\it New Horizons} spacecraft.  Its icy surface is probably covered with complex organics.  NASA image PIA19952.  }
   \label{new_horizon}
\end{center}
\end{figure}

Transneptunian objects (TNOs) are icy objects that are among the most primitive bodies in the Solar System.  
Spectroscopic observations of Pluto, one of the largest TNOs, have identified at least five different ices:  N$_2$, CH$_4$, CO, H$_2$O, and C$_2$H$_6$ \citep{cruik2015}.
Images of Pluto obtained from the recent {\it New Horizons} mission show a range of colored surface regions, suggesting the presence of complex organics embedded in water ice on the surface (Fig.~\ref{new_horizon}), as well as in liquid form in subsurface reservoirs \citep{pluto}.  The low albedo found in TNOs has also been suggested to be due to surface complex organics \citep{giri}.

\section{Organic synthesis in stars}

Recent infrared and millimeter-wave spectroscopic observations of evolved stars have shown that they are prolific molecular factories.  Over 80 gas-phase molecules have been detected through their rotational transitions in the circumstellar envelope of asymptotic giant branch (AGB) stars, evolved stars that are capable of synthesizing the element carbon through nuclear reactions \citep{ziurys}.  Minerals including silicates, silicon carbide and refractory oxides have been detected by infrared spectroscopy \citep{kwok2004}.  

\begin{figure}
\centering
\includegraphics[width=0.8\textwidth]{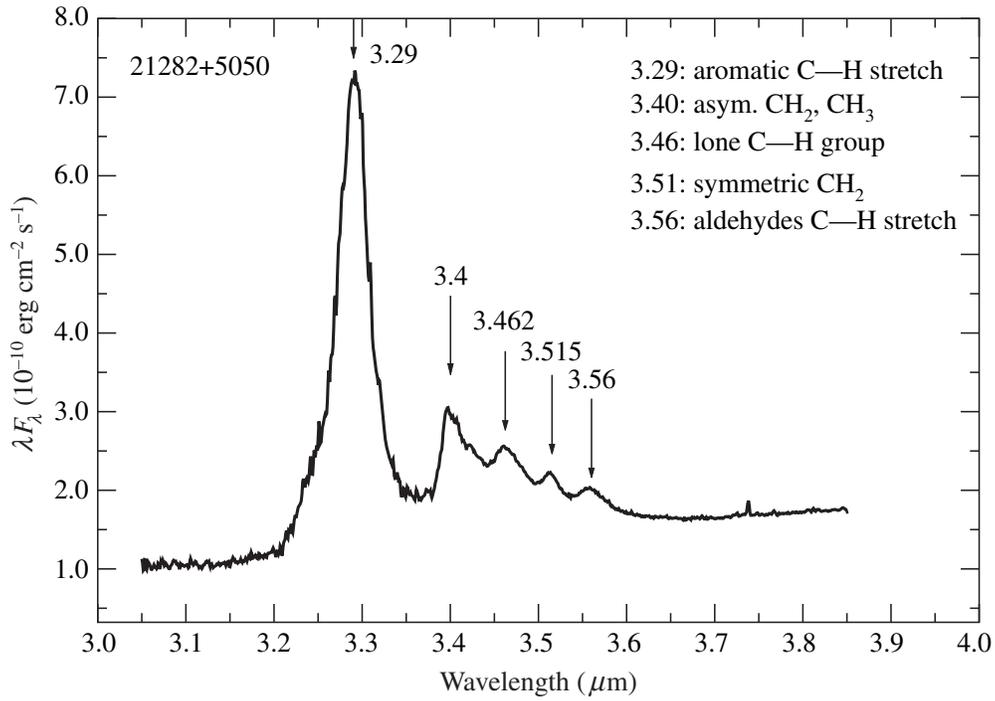}
\caption{Near-infrared spectrum of the young planetary nebula IRAS 21282+5050 obtained with the 10-m Keck Telescope, showing the symmetric and antisymmetric C--H stretching modes of methyl and methylene group in emission.   Figure adapted from \citet{hrivnak}.}
\label{keck}
\end{figure}

Most interestingly, complex organics can be seen forming in the descendents of AGB stars -- planetary nebulae.  
Comparison of the spectra of AGB stars, proto-planetary nebulae, and planetary nebulae shows progressive synthesis of molecules with increasing complexity.  Starting with simple molecules such as C$_2$, C$_3$, CN, chains (HCN, HC$_3$N, HC$_5$N), rings (C$_3$H$_2$), and aceylene (C$_2$H$_2$) are formed in the stellar winds of AGB stars.  In the proto-planetary nebulae stage,  di-acetylene, tri-acetylene, and the first aromatic molecule benzene are formed.  The first signs of aromatic and aliphatic structures also emerges in proto-planetary nebulae stage.

A family of unidentified infrared emission (UIE) bands at 3.3, 3.4, 6.2, 6.9, 7.7, 8.6, and 11.3 $\mu$m, first discovered in the young planetary nebula NGC 7027 \citep{russell}, have been suggested to be due to stretching and bending modes of aromatic and aliphatic hydrocarbons \citep{duley81}.  The 3.4 $\mu$m features observed in proto-planetary nebulae and young planetary nebulae \citep{hrivnak} (Figure \ref{keck}) are very similar to the 3.4 $\mu$m features seen in carbonaceous chondrites, comets \citep{keller2006}, Titan \citep{kim} and IDPs \citep{flynn03}.  These features sit on top of broad emission plateaus around 8 and 12 $\mu$m, which are probably due to superpositions of in-plane-bending and out-of-plane bending modes of a mixture of aliphatic groups \citep{kwok2001}.

\begin{figure}[h]
\begin{center}
\includegraphics[width=3.0in]{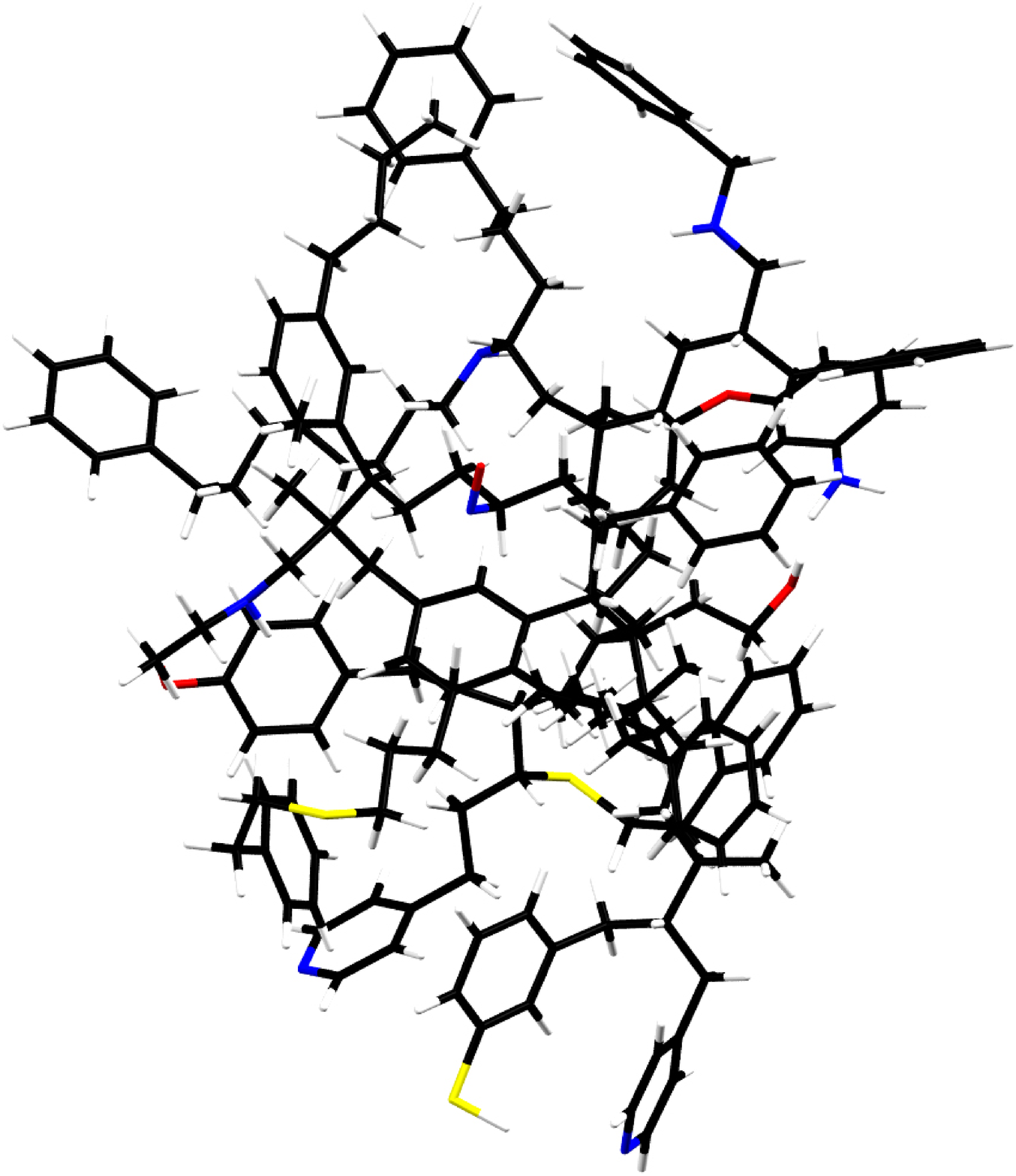} 
 \vspace*{-0.1 cm}
 \caption{The MAON structure is characterized by a highly disorganized arrangement of small units of aromatic rings linked by aliphatic chains.  This structure contains 169 C atoms (in black) and  225 H atoms (in white).  Impurities such as O (in red), N (in blue), and S (in yellow) are also present.  A typical MAON particle may consist of multiple structures similar to this one.}
   \label{maon}
\end{center}
\end{figure}

The UIE bands and their associated underlying broad emission plateaus have been interpreted as originating from nanoparticles of mixed aromatic-aliphatic structures \citep[MAON,][]{kz2011}. MAONs consist of multiple small islands of aromatic rings connected by aliphatic chains of varying lengths and orientations.  Beyond hydrogen and carbon, elements such as oxygen, sulphur, nitrogen are also present.  An example of the MAON structure is shown in Fig. \ref{maon}.  A typical particle may consist of hundreds or thousands of carbon atoms.  
These organic particles are formed in the circumstellar envelopes of evolved stars under low density ($<10^6$ cm$^{-3}$) conditions and on times scales of $\sim 10^3$ yr.  They are ejected via stellar winds into the interstellar medium.

\section{Delivery of stellar organics to the Solar System}

Although it is commonly believed  that all organics in the Solar System was synthesized  in situ after the formation of the Solar System, the discovery of stellar synthesis of complex organics raises the possibility that the primordial solar nebula may have been enriched by stellar ejecta.  MAON-like particles are not easily destroyed by interstellar UV radiation or shocks  and can  survive their journeys across the Galaxy. Since almost all stars in the Galaxy will go through the planetary nebulae stage \citep{kwok2000}, among which approximately half are carbon rich and are capable of making complex organics, it is quite plausible that most of the star and planetary forming regions contain organics ejected by nearby stars.  

When exposed to UV and charged particle irradiation, remnants of stellar organics embedded in icy planets, TNOs, and comets, could lead to the formation of a variety of pre-biotic molecules including amino acids and nucleobases \citep{pluto}. The organics found in carbonaceous chondrites could be the result of such processes.  

\section{Primordial hydrocarbons on Earth?}

Although the inventory of organic carbon near the Earth's surface is well documented (Table 1), the amount of carbon in the Earth's mantle and core is much less well known \citep{marty}.  The presence of primordial hydrocarbon deep inside the Earth is intriguing, but gas-phase organic molecules such as methane would have a difficult time surviving planetary differentiation after accretion   \citep{sephton2013}.

The terrestrial planets were formed as the result of aggregation of planetesimals.  It is possible that the early Earth may have incorporated some of the primordial organics during its formation.  Macromolecular organic solids similar to MAONs  would have a better chance of  withstanding the thermal and dynamical evolution of the early Earth.
Under suitable temperature and pressure conditions, e.g. those in hydrothermal vents, organic components such as hydrocarbons, dicarboxylic acids, N-, O-, and S-containing aromatic compounds, as well as ammonia can be released from macromolecular compounds \citep{yabuta, pizz2011}.  These prebiotic materials could form the ingredients of the first steps to life in the early Earth.

\section{Conclusions}

Since the first evidence of extratresstrial organics found in meteorites, we have learned that organics are commonly found in Solar System objects, evolved stars, interstellar clouds, diffuse interstellar medium, and distant galaxies \citep{kwok2011, kwok2016}.  These organics are not breakdown products of life, but are synthesized bottom up from simple molecules.  
Although the presence of organic matter is widely observed throughout the Universe, evolved stars are the only sites that organic synthesis can be directly observed to be happening.  
Observations of evolved stars provide direct evidence that the synthesis of complex organics is fast and efficient, and the ubiquitous presence of organics suggests that abiotic synthesis of organic matter is at work across the Universe. 
The fact that organic matter on Earth is almost exclusively biological in origin (Table 1) is actually an anomaly given the wide presence of abiotic organic matter in the Universe.  The discoveries of complex organics on Mars and Enceladus are therefore not unexpected and cannot be construed as evidence for life.

A large number of organic molecules, including pre-biotic molecules such as amino acids and nucleic acids, are found in meteorites, and their exact pathway of synthesis is unclear.  One possible scenario is that the organics in the soluble component of meteorites are processed products of IOM, which are related to MAONs ejected by stars.  In this picture, IOM is stellar in origin but the other organic molecules are produced within the Solar System on the surfaces of comets, TNOs, and planetary satellites.

The Earth may have inherited stellar organics that are either embedded in the interior of the primordial Earth \citep{kwok2017}, or later delivered by external bombardments of comets and asteroids \citep{chyba}.
Although we do not know the exact mechanism of how life originated on Earth from non-living matter, the possible presence of primordial organics on Earth suggests that the ``primordial soup'' may be richer in content than previously believed, which would have made the emergence of life easier \citep{ehrenfreund, kwok2009}.  
The likely connections between stellar and Solar System organics, the possible presence of primordial organics in the early Earth, and the potential role that they may play in the origin of life, are fascinating topics for further investigations.

\acknowledgments
I am grateful to Prof. Wing Ip and the scientific organizing committee of inviting me to give this talk, and to Prof. Kwing Chan and the local organizing committee for their hospitality.
This work is supported by a grant from the Natural Science and Engineering Research Council of Canada.


\clearpage

\end{document}